\documentclass[aps,pra,reprint,showpacs,amsmath,amssymb,10pt]{revtex4-1}
\usepackage{graphicx}
\usepackage{bm}
\usepackage{slashed}
\usepackage{color}

\begin{document}

\title{Coherent combs of anti-matter from nonlinear electron-positron pair creation}

\author{K. Krajewska}
\email[E-mail address:\;]{Katarzyna.Krajewska@fuw.edu.pl}
\author{J. Z. Kami\'nski}
\affiliation{Institute of Theoretical Physics, Faculty of Physics, University of Warsaw, Pasteura 5,
02-093 Warszawa, Poland}
\date{\today}
\begin{abstract}
Electron-positron pair creation in collisions of a modulated laser pulse with a high-energy photon (nonlinear Breit-Wheeler process) 
is studied by means of strong-field quantum electrodynamics. It is shown that the driving pulse modulations lead to 
appearance of comb structures in the energy spectra of produced positrons (electrons). It is demonstrated that these combs result from a coherent enhancement
of probability amplitudes of pair creation from different modulations of the laser pulse. Thus, resembling the Young-double slit experiment for
anti-matter (matter) waves.
\end{abstract}

\pacs{12.20.Ds, 12.90.+b, 42.55.Vc, 13.40.−f}
\maketitle

\maketitle

\section{Introduction}
\label{intro}

Production of relativistic charged particle beams is an important area of research, due to its various applications
in nuclear and particle physics, in plasma physics, and laboratory astrophysics. In recent years, with the rapid development 
of high-power laser technology, this field has been largely focused on laser-based techniques leading to the production and 
acceleration of particle beams. It has been demonstrated that laser-plasma accelerators can produce electron 
beams with energies around 1GeV~\cite{Kneip,Leemans,Hafz,Lundh,Wang}. One should mention that laser-driven high-energy positron
beams are much harder to generate. The first laser-driven positron beams have been achieved with energies less than 5MeV~\cite{Gahn1}.
The following experiments reached positron energies up to 20MeV~\cite{Gahn2,Chen1,Chen2}.
Most recently, positron beams with energies up to 120MeV have been produced in an all-optical setup experiment
carried out at the Hercules laser facility~\cite{Michigan1,Michigan2}.

The experiment at Hercules~\cite{Michigan1,Michigan2} has been based on a cascading mechanism initiated by the propagation of a laser-accelerated, 
relativistic electron beam through high-Z solid targets. The positrons inside the solid were mainly generated via either a trident process, 
in which electron-positron pair production is mediated by a virtual photon in the electron field, or via a two-step process where the electron first
emits a bremsstrahlung photon, which then produces an electron-positron pair via the Bethe-Heitler process. Another possible scenario of
producing electron-positron pairs is {\it the nonlinear Breit-Wheeler process}, which is the main topic of this paper. For recent reviews
concerning electron-positron pair creation, we refer the reader to~\cite{Ehlotzky,Ruffini,Ros1,HeidelbergReview,Ros2}.

In the original paper by Breit and Wheeler~\cite{Breit}, a collision of two photons which combined their energies
to produce an electron-positron pair was considered. This idea was further generalized by replacing one of the photons by 
a strong laser field~\cite{Reiss,Nikishov,Narozhny}. In these pioneering studies 
and in the following studies as well~\cite{Ivanov,Ivan,Denisenko}, the laser field was treated
as a monochromatic plane wave. The process in a weakly nonmonochromatic laser field was investigated by Narozhny and 
Fofanov in~\cite{Nar}. It is important to note that the laser-induced Breit-Wheeler process was successfully tested in a series of experiments 
carried out at SLAC~\cite{Bula,Bamber}.

The recent progress of ultra-intense lasers have stimulated even more advanced theoretical studies of the Breit-Wheeler process, which
involve pulsed laser fields~\cite{Heinzl,Titov1,Nousch,KasiaBW,Titov2,Kasiaproc}.
Different aspects of the process have been studied in these papers, including the dependence of probability distributions of created pairs
on intensity, shape, duration, and carrier-envelope phase of the driving pulse, as well as spin effects. Here,
we demonstrate for the first time how, by means of the laser-stimulated Breit-Wheeler process, one can generate positron energy combs.
For the parameters considered in this paper, we demonstrate a possibility of generating positron combs in the
energy region of tens of GeV. Note that GeV positrons were produced in SLAC experiment~\cite{Bula,Bamber}.
This required the generation of a high-energy photon, which then collided with a strong laser beam.
A 29GeV photon was produced by scattering 46.6GeV electrons off a 527nm laser light 
with intensity of 10$^{19}$W/cm$^2$ (for other parameters, see~\cite{Bula,Bamber}). The same laser beam was used in the following step, i.e., in the nonlinear Breit-Wheeler process.
Note that the parameters used throughout this paper are in the regime accessible to the SLAC experiment.

The organization of this paper is as follows. In Section~\ref{theory}, we present the theoretical formulation of the nonlinear Breit-Wheeler process which is along the lines
presented in our previous paper~\cite{KasiaBW}. In Section~\ref{shape},
we introduce the laser pulse shape for which numerical calculations are performed. The latter are illustrated
in Section~\ref{numerics}. In Section~\ref{diffraction}, we provide a theoretical explanation of comb-like structures
in the positron energy spectra, observed in Section~\ref{numerics}. A brief summary of our results is given in Section~\ref{conclusions}.

In the theoretical formulas below we keep $\hbar=1$, however the numerical results are presented in relativistic units
such that $\hbar=c=m_{\rm e}=1$ where $m_{\rm e}$ is the electron mass.

\section{Theory}
\label{theory}

The $S$-matrix amplitude for the Breit-Wheeler pair creation process, 
$\gamma_{\bm{K}\sigma}\rightarrow e^-_{\bm{p}_{{\rm e}^-}\lambda_{{\rm e}^-}}+e^+_{\bm{p}_{{\rm e}^+}\lambda_{{\rm e}^+}}$, 
with electron and positron momenta and spin polarizations 
$\bm{p}_{{\rm e}^-}\lambda_{{\rm e}^-}$ and $\bm{p}_{{\rm e}^+}\lambda_{{\rm e}^+}$, 
respectively, equals~\cite{KasiaBW}
\begin{equation}
{\cal A}=
-\mathrm{i}e\int \mathrm{d}^{4}{x}\, j^{(+-)}_{\bm{p}_{{\rm e}^-}\lambda_{{\rm e}^-},
\bm{p}_{{\rm e}^+}\lambda_{{\rm e}^+}}(x)\cdot A^{(+)}_{\bm{K}\sigma}(x), \label{BWAmplitude}
\end{equation}
where $\bm{K}\sigma$ denotes the initial nonlaser photon momentum and polarization. Here,
\begin{equation}
A^{(+)}_{\bm{K}\sigma}(x)=\sqrt{\frac{1}{2\varepsilon_0\omega_{\bm{K}}V}} 
\,\varepsilon_{\bm{K}\sigma}\mathrm{e}^{-\mathrm{i}K\cdot x},
\label{per}
\end{equation}
where $V$ is the quantization volume, $\varepsilon_0$ is the vacuum electric permittivity,
$\omega_{\bm{K}}=cK^0=c|\bm{K}|$ ($K\cdot K=0$), and $\varepsilon_{\bm{K}\sigma}=(0,\bm{\varepsilon}_{\bm{K}\sigma})$ 
is the linear polarization four-vector satisfying the conditions,
\begin{equation}
K\cdot\varepsilon_{\bm{K}\sigma}=0,\quad \varepsilon_{\bm{K}\sigma}\cdot\varepsilon_{\bm{K}\sigma'}=-\delta_{\sigma\sigma'},
\label{polargamma}
\end{equation}
for $\sigma,\sigma'=1,2$. The matrix element of the pair current operator is defined as
\begin{equation}
[j^{(+-)}_{\bm{p}_{{\rm e}^-}\lambda_{{\rm e}^-},\bm{p}_{{\rm e}^+}\lambda_{{\rm e}^+}}(x)]^{\nu}
=\bar{\psi}^{(+)}_{\bm{p}_{{\rm e}^-}\lambda_{{\rm e}^-}}(x)\gamma^\nu \psi^{(-)}_{\bm{p}_{{\rm e}^+}\lambda_{{\rm e}^+}}(x).
\end{equation}
Moreover, $\psi^{(\beta)}_{\bm{p}\lambda}(x)$ (with $\beta=+1$ for electron and $\beta=-1$ for positron) 
is the Volkov solution of the Dirac equation coupled to the electromagnetic field~\cite{Volkov,KK}
\begin{equation}
\psi^{(\beta)}_{\bm{p}\lambda}(x)=\sqrt{\frac{m_{\mathrm{e}}c^2}{VE_{\bm{p}}}}\Bigl(1-\beta\frac{e}{2k\cdot p}\slashed{A}\slashed{k}\Bigr)
u^{(\beta)}_{\bm{p}\lambda}\mathrm{e}^{-\mathrm{i} \beta S_p^{(\beta)}(x)} , \label{Volk}
\end{equation}
with the phase:
\begin{equation}
S_p^{(\beta)}(x)=p\cdot x+\int_{-\infty}^{k\cdot x} \Bigl[\beta\frac{ e A(\phi )\cdot p}{k\cdot p}
-\frac{e^2A^{2}(\phi )}{2k\cdot p}\Bigr]{\rm d}\phi .
\label{bbb}
\end{equation}
Here, $E_{\bm{p}}=cp^0\geq m_\mathrm{e}c^2$, $p=(p^0,\bm{p})$, $p\cdot p=(m_{\mathrm{e}}c)^2$, and $u^{(\beta)}_{\bm{p}\lambda}$ 
are the free-electron (positron) bispinors normalized such that
\begin{equation}
\bar{u}^{(\beta)}_{\bm{p}\lambda}u^{(\beta')}_{\bm{p}\lambda'}=\beta\delta_{\beta\beta'}\delta_{\lambda\lambda'}.
\end{equation}
The four-vector potential $A(k\cdot x)$ in Eq.~\eqref{Volk} represents a laser field, for which $k\cdot A(k\cdot x)=0$ and $k\cdot k=0$. 
Henceforth, we consider the Coulomb gauge for the radiation field which 
means that the four-vector $A(k\cdot x)$ has a vanishing zero component and that the 
electric and magnetic fields are equal to
\begin{eqnarray}
\bm{\mathcal{E}}(k\cdot x)&=-\partial_t \bm{A}(k\cdot x)= -ck^0 \bm{A}'(k\cdot x), \label{electric} \\
\bm{\mathcal{B}}(k\cdot x)&=\bm{\nabla}\times \bm{A}(k\cdot x)= -\bm{k}\times \bm{A}'(k\cdot x), \label{magnetic} 
\end{eqnarray}
where \textit{'prime'} means the derivative with respect to $k\cdot x$. We define the laser pulse such that
\begin{equation}
A(k\cdot x)=0 \quad \textrm{for} \quad k\cdot x < 0\quad \textrm{and}\quad k\cdot x > 2\pi.
\label{LaserVectorPotentialPulse}
\end{equation}
Note that with this definition, the condition imposed on the electric field generated by lasers:
\begin{equation}
\int_{-\infty}^{\infty}\bm{\mathcal{E}}(ck^0t-\bm{k}\cdot\bm{r})\mathrm{d}t=0, \label{LaserCondition1}
\end{equation}
is satisfied. We can also interpret the momentum $p$, which labels the Volkov
solution~\eqref{Volk}, as an asymptotic momentum of a free positron (electron).

Consider a laser pulse that propagates in the direction determined
by the unit vector ${\bm n}$, and that lasts for time $T_{\rm p}$. Its fundamental frequency 
is $\omega=2\pi/T_{\rm p}$ whereas its wave four-vector is $k=k^0(1,{\bm n})$, with $k^0=\omega/c$. 
The laser pulse is characterized by the following four-vector potential,
\begin{equation}
A(k\cdot x)=A_0\bigl[\varepsilon_1 f_1(k\cdot x)+\varepsilon_2 f_2(k\cdot x)\bigr], \label{LaserVectorPotential}
\end{equation}
where two real four-vectors $\varepsilon_i$ describe two linear polarizations and satisfy the relations,
\begin{equation}
\varepsilon_i^2=-1,\quad \varepsilon_1\cdot\varepsilon_2=0,\quad k\cdot\varepsilon_i=0 . \label{polarlaser}
\end{equation}
As explained in Ref.~\cite{KasiaBW},
such a choice of the polarization four-vector $\varepsilon$ as a spacelike vector (the same concerns
also $\varepsilon_{{\bm K}\sigma}$) is justified by gauge invariance of the probability amplitude
\eqref{BWAmplitude}. Keeping this in mind, we represent this amplitude such that
\begin{equation}
{\cal A}=\mathrm{i}\sqrt{\frac{2\pi\alpha c(m_{\mathrm{e}}c^2)^2}{E_{\bm{p}_{{\rm e}^-}}E_{\bm{p}_{{\rm e}^+}}\omega_{\bm{K}}V^3}}\, \mathcal{A}_{\mathrm{BW}}, \label{ct1}
\end{equation}
where $\alpha=e^2/(4\pi\varepsilon_0 c)$ is the fine-structure constant and
\begin{equation}
{\cal A}_{\mathrm{BW}}=\int{\rm d}^4x{\rm e}^{-\mathrm{i}(K\cdot x -S^{(+)}_{{p}_{{\rm e}^-}}(x)-S^{(-)}_{{p}_{{\rm e}^+}}(x))}F(k\cdot x), \label{ct2}
\end{equation}
with
\begin{eqnarray}
F(k\cdot x)&=&\bar{u}^{(+)}_{\bm{p}_{{\rm e}^-}\lambda_{{\rm e}^-}}\Bigl(1-\frac{\mu m_\mathrm{e}c}{2k\cdot p_{{\rm e}^-}}\bigl[f_1(k\cdot x)\slashed{\varepsilon}_1\slashed{k}\nonumber\\
&+&f_2(k\cdot x)\slashed{\varepsilon}_2\slashed{k}\bigr]\Bigr) \label{Ffunction} \slashed{\varepsilon}_{\bm{K}\sigma}\Bigl(1-\frac{\mu m_\mathrm{e}c}{2k\cdot p_{{\rm e}^+}} \bigl[f_1(k\cdot x)\slashed{\varepsilon}_1\slashed{k}\nonumber\\
&+&f_2(k\cdot x)\slashed{\varepsilon}_2\slashed{k}\bigr]\Bigr) u^{(-)}_{\bm{p}_{{\rm e}^+}\lambda_{{\rm e}^+}}. 
\end{eqnarray}
Here, we introduce an invariant laser strength parameter, $\displaystyle\mu=\frac{|eA_0|}{m_{\mathrm{e}}c}$. 
We further rewrite Eq.~\eqref{ct2} such that
\begin{eqnarray}
{\cal A}_{\mathrm{BW}}&=&\int{\rm d}^4x\exp\Bigl(-{\rm i}Q\cdot x-{\rm i}\int_{-\infty}^{k\cdot x}{\rm d}\phi\,h(\phi)\Bigr)\nonumber\\
&\times&\Bigl[F_0+F^{(\rm reg)}(k\cdot x)\Bigr],
\label{ampl}
\end{eqnarray}
where
\begin{eqnarray}
Q&=&K-p_{{\rm e}^-}-p_{{\rm e}^+},\label{QQ}\\
h(\phi)&=&a_1 f_1(\phi)+ a_2 f_2(\phi)+b[f^2_1(\phi)+f^2_2(\phi)] ,\label{hh}
\end{eqnarray}
with
\begin{eqnarray}
a_1&=& -\mu m_{\mathrm{e}}c\Bigl(\frac{\varepsilon_1\cdot p_{{\rm e}^+}}{k\cdot p_{{\rm e}^+}} - \frac{\varepsilon_1\cdot p_{{\rm e}^-}}{k\cdot p_{{\rm e}^-}}\Bigr), \nonumber \\
a_2&=& -\mu m_{\mathrm{e}}c\Bigl(\frac{\varepsilon_2\cdot p_{{\rm e}^+}}{k\cdot p_{{\rm e}^+}} - \frac{\varepsilon_2\cdot p_{{\rm e}^-}}{k\cdot p_{{\rm e}^-}}\Bigr), \nonumber \\
b&=& -\frac{1}{2}(\mu m_{\mathrm{e}}c)^2 \Bigl(\frac{1}{k\cdot p_{{\rm e}^+}} + \frac{1}{k\cdot p_{{\rm e}^-}}\Bigr).
\label{tilded}
\end{eqnarray}  
In addition, in Eq.~\eqref{ampl} we have introduced
\begin{equation}
F_0=\bar{u}^{(+)}_{\bm{p}_{{\rm e}^-}\lambda_{{\rm e}^-}}\slashed{\varepsilon}_{\bm{K}\sigma}u^{(-)}_{\bm{p}_{{\rm e}^+}\lambda_{{\rm e}^+}},
\end{equation}
and
\begin{equation}
F^{(\rm reg)}(k\cdot x)=F(k\cdot x)-F_0.
\end{equation}
For numerical purposes, in Eq.~\eqref{ampl} we perform the Boca-Florescu transformation~\cite{Compton,Boca}
which states that, as long as $Q^0\neq 0$,
\begin{eqnarray}
&&\int\mathrm{d}^{4}{x}\,\exp\Bigl(-\mathrm{i}Q\cdot x-\mathrm{i}\int_{-\infty}^{k\cdot x}\mathrm{d}\phi\,h(\phi)\Bigr) \\
&=&-\frac{k^0}{Q^0}\int\mathrm{d}^{4}{x} 
\,h(k\cdot x)\exp\Bigl(-\mathrm{i}Q\cdot x-\mathrm{i}\int_{-\infty}^{k\cdot x}\mathrm{d}\phi\,h(\phi)\Bigr) .\nonumber
\label{BocaFlorescu}
\end{eqnarray}
We showed in Ref.~\cite{KasiaBW} that, for the nonlinear Breit-Wheeler process, the applicability condition for the Boca-Florescu transformation is indeed satisfied.
Therefore, we obtain
\begin{eqnarray}
{\cal A}_{\mathrm{BW}}&=&\int{\rm d}^4x\exp\Bigl(-{\rm i}Q\cdot x-{\rm i}\int_{-\infty}^{k\cdot x}{\rm d}\phi\,h(\phi)\Bigr)\nonumber\\
&\times&\Bigl[F^{(\rm reg)}(k\cdot x)-\frac{k^0}{Q^0}F_0 h(k\cdot x)\Bigr].
\label{amplit}
\end{eqnarray}
Here the phase in the exponent function will be further modified. For this purpose we define the electron and the positron laser-dressed momenta,
\begin{eqnarray}
\bar{p}_{{\rm e}^\mp}=p_{{\rm e}^\mp} &\mp& \mu m_\mathrm{e} c\Bigl(\frac{\varepsilon_1\cdot p_{{\rm e}^\mp}}{k\cdot p_{{\rm e}^\mp}}\langle f_1\rangle 
+ \frac{\varepsilon_2\cdot p_{{\rm e}^\mp}}{k\cdot p_{{\rm e}^\mp}}\langle f_2\rangle\Bigr)k \nonumber \\ 
&+ & \frac{1}{2}(\mu m_\mathrm{e} c)^2\frac{\langle f_1^2\rangle+\langle f_2^2\rangle}{k\cdot p_{{\rm e}^\mp}}k ,  \label{cp1}
\end{eqnarray}
where
\begin{equation}
\langle f_i^j \rangle=\frac{1}{2\pi}\int\limits_0^{2\pi}\mathrm{d}(k\cdot x) [f_i(k\cdot x)]^j, 
\label{ct6}
\end{equation}
for $i, j=1,2$. In accordance with these definitions, we also introduce 
\begin{eqnarray}
\bar{Q}&=&K-\bar{p}_{{\rm e}^-}-\bar{p}_{{\rm e}^+},\\
G(k\cdot x)&=&\int_0^{k\cdot x}\mathrm{d}{\phi}\Bigl[ a_1 \bigl(f_1(\phi) -\langle f_1\rangle\bigr) +a_2\bigl(f_2(\phi)-\langle f_2\rangle\bigr) \nonumber\\
 &+&b\bigl(f_1^2(\phi)-\langle f_1^2\rangle + f_2^2(\phi)-\langle f_2^2\rangle\bigr) \Bigr],  \label{cp4}
\end{eqnarray}
where $G(0)=G(2\pi)=0$. Thus, the probability amplitude of the nonlinear Breit-Wheeler process~\eqref{amplit} can be written as
\begin{eqnarray}
{\cal A}_{\mathrm{BW}}&=&\int{\rm d}^4x\,{\rm e}^{-{\rm i}\bar{Q}\cdot x-{\rm i}G(k\cdot x)}\nonumber\\
&\times&\Bigl[F^{(\rm reg)}(k\cdot x)-\frac{k^0}{Q^0}F_0 h(k\cdot x)\Bigr].
\label{amplit2}
\end{eqnarray}
Because the functions $f_i(k\cdot x)$ are non-zero except for the region from 0 to $2\pi$, we can introduce the following basis expansion in the interval $[0,2\pi]$,
\begin{equation}
[f_1(k\cdot x)]^n[f_2(k\cdot x)]^m \mathrm{e}^{-\mathrm{i}G(k\cdot x)}
=\sum_{N=-\infty}^\infty G^{(n,m)}_N \mathrm{e}^{-\mathrm{i}Nk\cdot x}, \label{ct7}
\end{equation}
with integer $N$. This leads to
\begin{equation}
\mathcal{A}_{\mathrm{BW}}=\sum_N D_N \int\mathrm{d}^{4}{x}\,\mathrm{e}^{-\mathrm{i}(\bar{Q}+Nk)\cdot x}, \label{ct8}
\end{equation}
where the coefficients $D_N$ equal
\begin{widetext}
\begin{align}
D_N=&-\frac{1}{2}\mu m_\mathrm{e} c\Bigl[\Bigl( \frac{1}{k\cdot p_{{\rm e}^+}} \bar{u}^{(+)}_{\bm{p}_{{\rm e}^-}\lambda_{{\rm e}^-}}\slashed{\varepsilon}_{\bm{K}\sigma}  \slashed{\varepsilon}_1\slashed{k} u^{(-)}_{\bm{p}_{{\rm e}^+}\lambda_{{\rm e}^+}}
 +\frac{1}{k\cdot p_{{\rm e}^-}} \bar{u}^{(+)}_{\bm{p}_{{\rm e}^-}\lambda_{{\rm e}^-}} \slashed{\varepsilon}_1\slashed{k}\slashed{\varepsilon}_{\bm{K}\sigma}  u^{(-)}_{\bm{p}_{{\rm e}^+}\lambda_{{\rm e}^+}}
 \nonumber \\ &\qquad
 +\frac{2k^0}{Q^0}\bigl(\frac{\varepsilon_1\cdot p_{{\rm e}^+}}{k\cdot p_{{\rm e}^+}} - \frac{\varepsilon_1\cdot p_{{\rm e}^-}}{k\cdot p_{{\rm e}^-}}\bigr)\bar{u}^{(+)}_{\bm{p}_{{\rm e}^-}\lambda_{{\rm e}^-}} \slashed{\varepsilon}_{\bm{K}\sigma}  u^{(-)}_{\bm{p}_{{\rm e}^+}\lambda_{{\rm e}^+}}
\Bigr)G^{(1,0)}_N  \nonumber \\
 &\qquad +\Bigl( \frac{1}{k\cdot p_{{\rm e}^+}} \bar{u}^{(+)}_{\bm{p}_{{\rm e}^-}\lambda_{{\rm e}^-}} \slashed{\varepsilon}_{\bm{K}\sigma} \slashed{\varepsilon}_2\slashed{k} u^{(-)}_{\bm{p}_{{\rm e}^+}\lambda_{{\rm e}^+}}  
+\frac{1}{k\cdot p_{{\rm e}^-}} \bar{u}^{(+)}_{\bm{p}_{{\rm e}^-}\lambda_{{\rm e}^-}} \slashed{\varepsilon}_2\slashed{k}\slashed{\varepsilon}_{\bm{K}\sigma}  u^{(-)}_{\bm{p}_{{\rm e}^+}\lambda_{{\rm e}^+}} \nonumber \\ &\qquad
+\frac{2k^0}{Q^0}\bigl(\frac{\varepsilon_2\cdot p_{{\rm e}^+}}{k\cdot p_{{\rm e}^+}} - \frac{\varepsilon_2\cdot p_{{\rm e}^-}}{k\cdot p_{{\rm e}^-}}\bigr)\bar{u}^{(+)}_{\bm{p}_{{\rm e}^-}\lambda_{{\rm e}^-}} \slashed{\varepsilon}_{\bm{K}\sigma}  u^{(-)}_{\bm{p}_{{\rm e}^+}\lambda_{{\rm e}^+}}\Bigr)G^{(0,1)}_N\Bigr]  \nonumber \\
 &+  \frac{(\mu m_\mathrm{e} c)^2}{4(k\cdot p_{{\rm e}^+})(k\cdot p_{{\rm e}^-})} \Bigl[\Bigl(\bar{u}^{(+)}_{\bm{p}_{{\rm e}^-}\lambda_{{\rm e}^-}} \slashed{\varepsilon}_1\slashed{k} \slashed{\varepsilon}_{\bm{K}\sigma} \slashed{\varepsilon}_1\slashed{k} u^{(-)}_{\bm{p}_{{\rm e}^+}\lambda_{{\rm e}^+}}
 +\frac{2k^0}{Q^0}(k\cdot p_{{\rm e}^-}+k\cdot p_{{\rm e}^+})\bar{u}^{(+)}_{\bm{p}_{{\rm e}^-}\lambda_{{\rm e}^-}} \slashed{\varepsilon}_{\bm{K}\sigma}  u^{(-)}_{\bm{p}_{{\rm e}^+}\lambda_{{\rm e}^+}}\Bigr)G^{(2,0)}_N   \nonumber \\ &\qquad
+\Bigl(\bar{u}^{(+)}_{\bm{p}_{{\rm e}^-}\lambda_{{\rm e}^-}} \slashed{\varepsilon}_2\slashed{k} \slashed{\varepsilon}_{\bm{K}\sigma} \slashed{\varepsilon}_2\slashed{k} u^{(-)}_{\bm{p}_{{\rm e}^+}\lambda_{{\rm e}^+}}
 +\frac{2k^0}{Q^0}(k\cdot p_{{\rm e}^-}+k\cdot p_{{\rm e}^+})\bar{u}^{(+)}_{\bm{p}_{{\rm e}^-}\lambda_{{\rm e}^-}} \slashed{\varepsilon}_{\bm{K}\sigma}  u^{(-)}_{\bm{p}_{{\rm e}^+}\lambda_{{\rm e}^+}}\Bigr)G^{(0,2)}_N  \nonumber \\
 & \qquad +\Bigl(\bar{u}^{(+)}_{\bm{p}_{{\rm e}^-}\lambda_{{\rm e}^-}} \slashed{\varepsilon}_1\slashed{k} \slashed{\varepsilon}_{\bm{K}\sigma} \slashed{\varepsilon}_2\slashed{k} u^{(-)}_{\bm{p}_{{\rm e}^+}\lambda_{{\rm e}^+}}
  +\bar{u}^{(+)}_{\bm{p}_{{\rm e}^-}\lambda_{{\rm e}^-}} \slashed{\varepsilon}_2\slashed{k} \slashed{\varepsilon}_{\bm{K}\sigma} \slashed{\varepsilon}_1\slashed{k} u^{(-)}_{\bm{p}_{{\rm e}^+}\lambda_{{\rm e}^+}}  \Bigr)G^{(1,1)}_N \Bigr]. 
\end{align}
\end{widetext}

In order to perform the four-fold integral in Eq.~\eqref{ct8}, we introduce the light-cone coordinates, $(x^-,x^+,{\bm x}^\perp)$, 
such that $x^-=x^0-{\bm n}\cdot {\bm x}, x^+=\frac{1}{2}(x^0+{\bm n}\cdot {\bm x})$,
and ${\bm x}^\perp={\bm x}-({\bm n}\cdot {\bm x}){\bm n}$ \cite{Compton}. The integrals over $x^+$ and ${\bm x}^\perp$ lead to the delta conservation conditions while
the remaining integral is limited to the finite region, $0\leq x^-\leq 2\pi/k^0$. We obtain that
\begin{equation}
\mathcal{A}_{\mathrm{BW}}=\sum_N (2\pi)^3\delta^{(1)}(P_N^-)\delta^{(2)}(\bm{P}_N^\bot)D_N\frac{1-\mathrm{e}^{-2\pi\mathrm{i}P_N^+/k^0}}{\mathrm{i}P_N^+} ,  \label{cp5}
\end{equation}
where 
\begin{equation}
P_N=\bar{Q}+Nk. \label{cp6}
\end{equation}
Finally, taking into account that $\bm{P}_N^\bot$ and $P_N^{-}$ do not depend explicitly on $N$, and
\begin{equation}
\int\mathrm{d}^{3}p_{{\rm e}^-}\,\delta^{(1)}(P_N^{-})\delta^{(2)}(\bm{P}_N^\bot)=\frac{k^0p_{{\rm e}^-}^0}{k\cdot p_{{\rm e}^-}} ,
\end{equation}
we derive that the differential probability distribution of positrons produced via Breit-Wheeler scenario involving a finite laser pulse is
\begin{equation}
\frac{\mathrm{d}^{3}{\mathsf{P}}}{\mathrm{d}{E_{\bm{p}_{{\rm e}^+}}}
\mathrm{d}^{2}{\Omega_{\bm{p}_{{\rm e}^+}}}} = \frac{\alpha (m_{\mathrm{e}}c)^2 |\bm{p}_{{\rm e}^+}|}{k^0\omega_{\bm{K}}(k\cdot p_{{\rm e}^-})}   
\Big|\sum_N D_N\frac{1-\mathrm{e}^{-2\pi\mathrm{i}P_N^0/k^0}}{2\pi{\rm i} P_N^0/k^0} \Big|^2 ,  \label{cp10}
\end{equation}
where the electron four-momentum $p_{{\rm e}^-}$ equals~\cite{KasiaBW}
\begin{align}
\bm{p}_{{\rm e}^-}^{\bot} &=\bm{w}^{\bot} , \\
p_{{\rm e}^-}^{\|} &= \frac{(m_{\mathrm{e}}c)^2+(\bm{w}^{\bot})^2-(w^-)^2}{2w^-} ,  \\
p_{{\rm e}^-}^0    &= \frac{(m_{\mathrm{e}}c)^2+(\bm{w}^{\bot})^2+(w^-)^2}{2w^-} ,
\end{align}
and $w=K-p_{{\rm e}^+}$. This set of equations follow from solving the momentum conservation conditions, $P_N^-=0$ and ${\bm P}_N^\perp={\bm 0}$.
One can check that these solutions satisfy the on-mass shell relation, $p_{{\rm e}^-}\cdot p_{{\rm e}^-}=(m_{\mathrm{e}}c)^2$.
Note also that the corresponding probability distribution for electrons can be obtained from~\eqref{cp10}
by interchanging labels ${\rm e}^+$ and ${\rm e}^-$.

As a final remark, we note that the expression under the modulus in Eq.~\eqref{cp10}, can be rewritten as
\begin{align}
\sum_N D_N&\frac{1-\mathrm{e}^{-2\pi\mathrm{i}P_N^0/k^0}}{2\pi{\rm i} P_N^0/k^0}=\nonumber\\
&{\rm e}^{{\rm i}\pi N_{\rm eff}}\sum_N (-1)^N D_N\,{\rm sinc}[\pi(N-N_{\rm eff})],
\label{something}
\end{align}
where ${\rm sinc}(x)=\sin(x)/x$, and where we define~\cite{KasiaBW}
\begin{equation}
 N_{\rm eff}=(\bar{p}_{{\rm e}^-}^0+\bar{p}_{{\rm e}^+}^0-K^0)/k^0\equiv-\bar{Q}^0/k^0.
\label{neff}
\end{equation}
The phase factor in~\eqref{something}, $\pi N_{\rm eff}$, can be represented as
\begin{equation}
\pi N_{\rm eff}=\frac{\frac{\omega_{\rm cut}}{\omega}}{1-\frac{\omega_{\rm cut}}{\omega_{\bm K}}}{\cal F},
\label{faza}
\end{equation}
where the {\it cutoff} frequency equals
\begin{equation}
\omega_{\rm cut}=c\frac{n\cdot p_{{\rm e}^+}}{n\cdot n_{\bm K}}.
\label{faza1}
\end{equation}
Also, we have introduced
\begin{equation}
{\cal F}=\pi\frac{n_{\bm K}\cdot p_{{\rm e}^+}}{n\cdot p_{{\rm e}^+}}\Bigl(1+{\cal F}_1+{\cal F}_2+{\cal F}_{\rm sq}\Bigr),
\end{equation}
where (for $i=1,2$)
\begin{align}
{\cal F}_i=&\mu m_{\rm e}c\,\langle f_i\rangle \biggl[\frac{(\varepsilon_i\cdot n_{\bm K})
(n\cdot n_{\bm K})}{(n\cdot p_{{\rm e}^+})(n_{\bm K}\cdot p_{{\rm e}^+})}-\frac{n_{\bm K}\cdot \varepsilon_i}{n_{\bm K}\cdot p_{{\rm e}^+}}\biggr],\\
{\cal F}_{\rm sq}=&\frac{1}{2}(\mu m_{\rm e}c)^2\bigl(\langle f_1\rangle^2+\langle f_2\rangle^2\bigr)\frac{n\cdot n_{\bm K}}{(n\cdot p_{{\rm e}^+})(n_{\bm K}\cdot p_{{\rm e}^+})}.
\end{align}
Taking into account the dependence on $p_{{\rm e}^+}^0$ of the phase~\eqref{faza}, where one has to account for Eq.~\eqref{faza1},
we expect that the energy distribution of pair creation will tend to zero when $\omega_{\rm cut}$ tends to $\omega_{\bm K}$. 
This imposes the bounds on the energy of produced positrons, $p_{{\rm e}^+}^0$. More specifically, pair production is allowed only if the condition $\omega_{\rm cut}<\omega_{\bm K}$ is met. 
This requirement is always satisfied if ${\bm n}\cdot{\bm n}_{{{\rm e}^+}}>0$, where ${\bm n}_{{{\rm e}^+}}$ denotes the propagation direction of the positron measured at the detector.
When ${\bm n}\cdot{\bm n}_{{{\rm e}^+}}<0$, the aforementioned requirement means that
\begin{gather}
(p_{{\rm e}^+}^0)^2\bigl[1-({\bm n}\cdot{\bm n}_{{{\rm e}^+}})^2\bigr]-2p_{{\rm e}^+}^0K^-+(K^-)^2\nonumber\\
+(m_{\rm e}c)^2\bigl({\bm n}\cdot{\bm n}_{{{\rm e}^+}}\bigr)^2> 0,
\label{ineq}
\end{gather}
and it has real solutions provided that 
\begin{equation}
K^-\geqslant m_{\rm e}c\,\sqrt{1-({\bm n}\cdot{\bm n}_{{{\rm e}^+}})^2}.
\end{equation}
Denoting as 
\begin{align}
 p_{{\rm e}^+}^{0,{\rm min}}=&\frac{K^-+{\bm n}\cdot{\bm n}_{{{\rm e}^+}}\sqrt{(K^-)^2-(m_{\rm e}c)^2\bigl[1-({\bm n}\cdot{\bm n}_{{{\rm e}^+}})^2\bigr]}}
{1-({\bm n}\cdot{\bm n}_{{{\rm e}^+}})^2},\\
 p_{{\rm e}^+}^{0,{\rm max}}=&\frac{K^--{\bm n}\cdot{\bm n}_{{{\rm e}^+}}\sqrt{(K^-)^2-(m_{\rm e}c)^2\bigl[1-({\bm n}\cdot{\bm n}_{{{\rm e}^+}})^2\bigr]}}
{1-({\bm n}\cdot{\bm n}_{{{\rm e}^+}})^2},
\end{align}
we find out that~\eqref{ineq} is satisfied for $m_{\rm e}c<p_{{\rm e}^+}^{0}<p_{{\rm e}^+}^{0,{\rm min}}$ or $p_{{\rm e}^+}^{0}>p_{{\rm e}^+}^{0,{\rm max}}$.
It follows from the condition $\omega_{\rm cut}<\omega_{\bm K}$ that when ${\bm n}\cdot{\bm n}_{{{\rm e}^+}}<0$ we have $K^->p_{{\rm e}^+}^0$.
One can check that while $p_{{\rm e}^+}^{0,{\rm min}}$ meets this requirement, $p_{{\rm e}^+}^{0,{\rm max}}$ does not. Hence, we 
arrive at the conclusion that a nonlinear Breit-Wheeler process produces relativistic positrons with energies 
\begin{equation}
m_{\rm e}c^2<cp_{{\rm e}^+}^{0}<cp_{{\rm e}^+}^{0,{\rm min}}. 
\end{equation}
Note that the respective bounds for positron final momenta have been derived in Ref.~\cite{KasiaBW} from purely kinematic arguments.
For this reason, the phase $\pi N_{\rm eff}$ introduced in~\eqref{faza} can be called a {\it kinematic phase} of probability amplitude of pair creation,
\begin{equation}
\Phi_{\rm BW}^{\rm kin}=\pi N_{\rm eff}.
\label{kinphase}
\end{equation}
Except of the kinematic phase, the probability amplitude of pair creation acquires also a {\it dynamic phase}, which can be defined based on 
Eq.~\eqref{something} as
\begin{equation}
\Phi_{\rm BW}^{\rm dyn}={\rm arg}\Bigl\{\sum_N (-1)^N D_N\,{\rm sinc}[\pi(N-N_{\rm eff})]\Bigr\}.
\label{dynphase}
\end{equation}
The latter can be calculated only numerically. Later on, we shall also relate to a global phase of Breit-Wheeler process,
\begin{equation}
\Phi_{\rm BW}=\Phi_{\rm BW}^{\rm kin}+\Phi_{\rm BW}^{\rm dyn},
\label{globalphase}
\end{equation}
when discussing our numerical results.

\section{Pulse shape}
\label{shape}

For numerical illustrations, we choose a linearly polarized laser pulse that is composed of $N_{\rm rep}$ identical subpulses.
This means that the four-vector potential~\eqref{LaserVectorPotential} describing such a pulse can be written as
\begin{equation}
 A(k\cdot x)=A_0N_{\rm rep}N_{\rm osc}\varepsilon f(k\cdot x),
\label{fourvector}
\end{equation}
where in Eq.~\eqref{LaserVectorPotential} we put $f_2(k\cdot x)=0$, and henceforth we keep $f_1(k\cdot x)\equiv f(k\cdot x)$ and $\varepsilon_1\equiv\varepsilon$.
Note that $\varepsilon=(0,{\bm\varepsilon})$ is a linear polarization four-vector, $A_0$ is related to the parameter $\mu$, and $N_{\rm osc}$ denotes the number of cycles in each subpulse.
We choose the following shape function
\begin{equation}
 f'(k\cdot x)=N_A\sin^2\Bigl(N_{\rm rep}\frac{k\cdot x}{2}\Bigr)\sin(N_{\rm rep}N_{\rm osc}k\cdot x),
\end{equation}
with the carrier frequency of the laser pulse, $\omega_{\rm L}=N_{\rm rep}N_{\rm osc}\omega$. In this case, the parameter
$N_{\rm eff}/(N_{\rm rep}N_{\rm osc})$, where $N_{\rm eff}$ is defined by Eq.~\eqref{neff}, has the meaning of a net number of 
photons exchanged with the laser field in order to create pairs. However, this interpretation makes sense only for relatively long laser pulses,
for which $\langle f_i\rangle$ ($i=1,2$) in Eq.~\eqref{cp1} tends to zero.
In the above definition of $f'(k\cdot x)$, the parameter $N_A$ is introduced such that
\begin{equation}
 \frac{1}{2\pi}\int_0^{2\pi}{\rm d}(k\cdot x) [f'(k\cdot x)]^2=\frac{1}{2},
\end{equation}
where the derivative is with respect to $k\cdot x$. As argued in Ref.~\cite{BH-no-spin}, with this condition the mean intensity
in the pulse is fixed. In the following, we also assume that the laser pulse propagates in the $z$-direction (${\bm n}={\bm e}_z$) 
and its polarization vector is ${\bm\varepsilon}={\bm e}_x$. In the considered configuration, the nonlaser photon is counterpropagating
(${\bm n}_{\bm K}=-{\bm e}_z$) and its polarization vector is either parallel, ${\bm\varepsilon}_{\bm K}={\bm e}_x$, or perpendicular, ${\bm\varepsilon}_{\bm K}=-{\bm e}_y$, 
to the polarization direction of the laser pulse.

\section{Numerical illustrations}
\label{numerics}

\begin{figure}
\centering
\includegraphics[width=8cm]{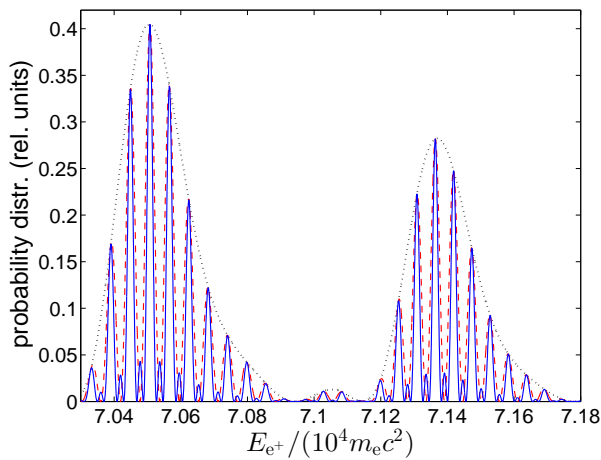}\\
\caption{The positron differential probability distribution (in rel. units) for laser-induced Breit-Wheeler pair creation 
calculated in the laboratory frame of reference. The driving laser pulse (characterized by parameters $\omega_{\mathrm{L}}=1.548\,\mathrm{eV}$, $\mu=1$, $N_{\mathrm{osc}}=16$,
and ${\bm\varepsilon}={\bm e}_x$) collides with a counterpropagating nonlaser photon of energy $10^5m_{\mathrm{e}}c^2$ and a colinear polarization. The positrons are created 
in the direction such that $\theta_{\mathrm{e}^+}=0.99999\pi$ and $\varphi_{\mathrm{e}^+}=0$. The results are presented for different numbers of modulated
subpulses, $N_{\rm rep}$. While the black dotted line corresponds to $N_{\rm rep}=1$, the red dashed line is for $N_{\rm rep}=2$, whereas the blue solid line is
for $N_{\rm rep}=3$. The results have been summed up with respect to the
spin degrees of freedom of produced particles. The probability distributions are divided by $N_{\mathrm{rep}}^2$.
\label{bwcomb20130322par}}
\end{figure}

\begin{figure}
\centering
\includegraphics[width=8cm]{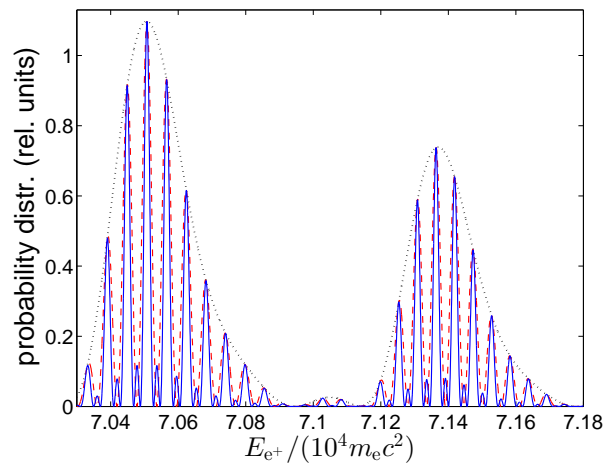}\\
\caption{The same as in Fig.~\ref{bwcomb20130322par} but a nonlaser photon is polarized opposite to the $y$-direction. 
\label{bwcomb20130322per}}
\end{figure}

In Figs.~\ref{bwcomb20130322par} and~\ref{bwcomb20130322per}, we demonstrate the positron energy spectra resulting from a head-on collision of a laser pulse with a high-energy photon,
calculated in the laboratory frame of reference. Unless otherwise stated, the parameters of the driving pulse are such that $\omega_{\mathrm{L}}=1.548\,\mathrm{eV}$, $\mu=1$, 
and $N_{\mathrm{osc}}=16$, whereas the energy of the colliding photon is $10^5m_{\mathrm{e}}c^2$. Note that for a Ti:sapphire laser light, $\mu=1$ corresponds roughly to the intensity 
of $10^{18}$ W/cm$^2$, which is currently attainable in experimental setups. In addition, GeV nonlaser photons can be generated, for instance,
via nonlinear Compton scattering, as it was demonstrated in the SLAC experiment~\cite{Bula}. We choose for the positron to be 
detected nearly along the propagation direction of the high-energy photon, i.e., for $\theta_{\mathrm{e}^+}=0.99999\pi$ and $\varphi_{\mathrm{e}^+}=0$.
In the figures, we show the results for a single ($N_{\rm rep}=1$; black dotted line), a double ($N_{\rm rep}=2$; dashed red line), 
and a triple ($N_{\rm rep}=3$; solid blue line) driving pulse. The results for different $N_{\rm rep}$ are scaled with $N_{\rm rep}^2$. 
For comparison, we show the spectra calculated for parallel (Fig.~\ref{bwcomb20130322par}) and perpendicular (Fig.~\ref{bwcomb20130322per})
polarizations of the laser pulse and the nonlaser photon. While the spectra differ in magnitude, their remaining properties are the same.
For this reason, we shall refer to Fig.~\ref{bwcomb20130322par} only.


\begin{figure}
\centering
\includegraphics[width=8.5cm]{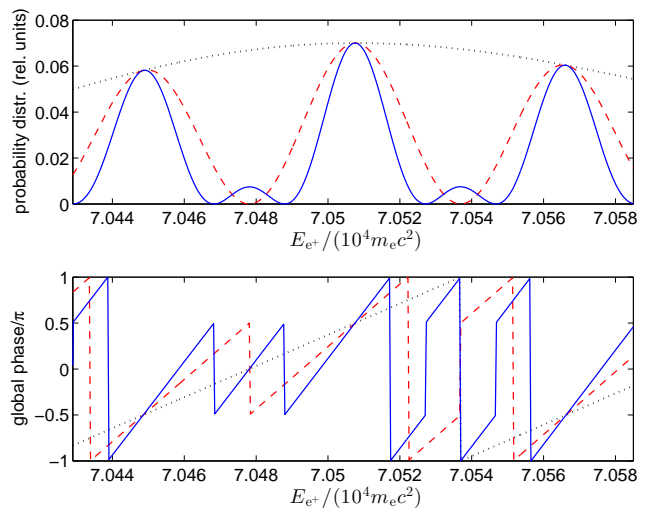}\\
\caption{The global phase~\eqref{globalphase} (lower panel) as a function of the positron energy for the parameters related to Fig.~\ref{bwcomb20130322par}.
For reference, in the upper panel a small portion of Fig.~\ref{bwcomb20130322par} is presented. At the comb maxima, the global phase takes the same value,
independent of $N_{\rm rep}$, and it varies by $\pi$ between two consecutive maxima. The same color coding is used as in Fig.~\ref{bwcomb20130322par}.
\label{bwc11}}
\end{figure}

It follows from Fig.~\ref{bwcomb20130322par} that, while the spectra for an unmodulated laser pulse ($N_{\rm rep}=1$) form broad peak structures, the spectra for
modulated pulses ($N_{\rm rep}=2,3$) show a clearly distinct structure composed of very narrow peaks. In the latter case, at certain positron energies 
(which are independent of $N_{\rm rep}$), we observe pronounced enhancements of the spectra. These high peaks are accompanied by smaller maxima. For a given $N_{\rm rep}$, there are always 
$(N_{\rm rep}-2)$ small maxima in-between the large ones. This typical behavior is also observed for bigger $N_{\rm rep}$.
Interestingly, the heights of the peaks are the same when scaled with $N_{\rm rep}^2$. These coherence properties are confirmed
by calculations of the global phase of probability amplitude~\eqref{globalphase} for each $N_{\rm rep}$, as shown in Fig.~\ref{bwc11}. For convenience, we keep the same
color scheme as in Fig.~\ref{bwcomb20130322par}. We observe that the global phase, $\Phi_{\rm BW}$, for each $N_{\rm rep}$ takes the same value
at each subsequent large maximum. Also, it changes by $\pi$ between two consecutive maxima. 

\begin{figure}
\centering
\includegraphics[width=8cm]{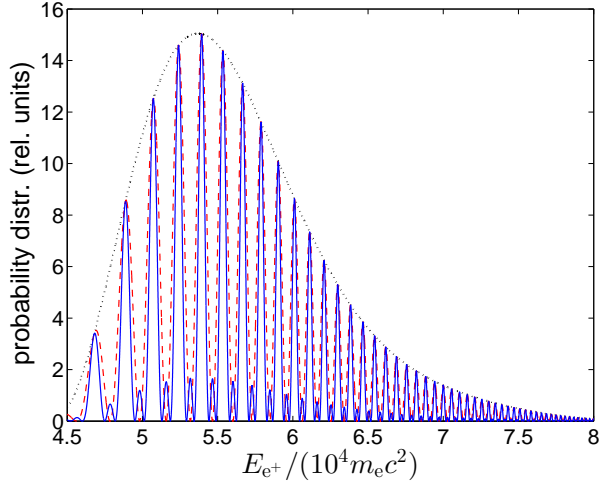}\\
\caption{The same as in Fig.~\ref{bwcomb20130322par} but for two-cycle subpulses ($N_{\rm osc}=2$). The spectra are presented 
over a large energy interval of positrons, reaching 40GeV.
\label{ssll2}}
\end{figure}

The same qualitative behavior of positron energy spectra is observed for other parameters
of the laser beam and the colliding high-energy photon. See, for instance, Fig.~\ref{ssll2} which presents the results for the case
when each subpulse is composed of two cycles ($N_{\rm osc}=2$). When comparing these results with the results of Fig.~\ref{bwcomb20130322par},
we note that less interference for $N_{\rm osc}=2$ leads to much broader peaks in the positron energy spectra. At the same time,
the probability distributions of created positrons increase in magnitude with decreasing duration of a single modulation, $N_{\rm osc}$. More precisely,
the probability distributions for $N_{\rm osc}=2$ are by an order of magnitude larger than those for $N_{\rm osc}=16$. We recall that 
the incident pulse has been defined such that it carries the same mean intensity, independent of its duration. If we keep the mean intensity
fixed, the maximum value of the electric and magnetic fields must be effectively increased when decreasing
the pulse duration. Consequently, an overall enhancement of the probability of pair production induced by shorter laser
pulses is observed. The same conclusion was reached in~\cite{KasiaBW} for unmodulated driving pulses. At this point, we also note
that while the peaks in Fig.~\ref{bwcomb20130322par} are equally spaced, the peaks in Fig.~\ref{ssll2} are not. As we shall explain later on, 
in general the comb structures are distributed nonuniformly on the positron energy scale. However, one can choose the finite energy intervals with 
uniformly spaced combs.

\section{Diffraction-type formula for nonlinear Breit-Wheeler process}
\label{diffraction}

How, by means of nonlinear Breit-Wheeler process, can one generate a comb in the energy domain of positrons 
using a laser pulse consisting of $N_{\rm rep}$ identical subpulses? To answer this question, we note that 
the four-vector potential~\eqref{LaserVectorPotential} (and, hence, also~\eqref{fourvector}) is a modulated
function of $x^-$, with the modulation period $\ell=2\pi/(k^0N_{\rm rep})$. Going back to Eq.~\eqref{amplit2}, we represent it as
\begin{eqnarray}
{\cal A}_{\mathrm{BW}}&=&(2\pi)^3\delta^{(1)}(Q^-)\delta^{(2)}({\bm Q}^\perp)\int_0^{N_{\rm rep}\ell }{\rm d}x^-\label{wwww}\\
&\times&{\rm e}^{-\mathrm{i}\bar{Q}^0 x^--\mathrm{i}G(k^0x^-)}\Bigl[F^{(\rm reg)}(k^0x^-)-\frac{k^0}{Q^0}F_0h(k^0x^-)\Bigr]. \nonumber
\end{eqnarray}
Due to the relation~\eqref{cp6}, the above conservation conditions are essentially the same as those introduced in Eq.~\eqref{cp5}. Moreover, the way we had defined
the incident pulse made it possible to restrict the integration over $x^-$ to a finite interval from 0 to $N_{\rm rep}\ell$. 
Now, we divide it into $N_{\rm rep}$ intervals, each of length $\ell$. As a result, the integral in Eq.~\eqref{wwww} can be represented as a sum of integrals,
\begin{align}
{\cal I}_n=\int_{(n-1)\ell}^{n\ell}&{\rm d}x^-{\rm e}^{-\mathrm{i}\bar{Q}^0 x^--\mathrm{i}G(k^0x^-)}\nonumber\\
&\times\Bigl[F^{(\rm reg)}(k^0x^-)-\frac{k^0}{Q^0}F_0h(k^0x^-)\Bigr],
\end{align}
where $n=1, 2,..., N_{\rm rep}$. One can easily derive that ${\cal I}_n={\rm e}^{-{\rm i}\bar{Q}^0(n-1)\ell}\,{\cal I}_1$, leading to 
the following formula for the probability amplitude of Breit-Wheeler process,
\begin{align}
{\cal A}_{\mathrm{BW}}&=(2\pi)^3\delta^{(1)}(Q^-)\delta^{(2)}({\bm Q}^\perp)\nonumber\\
&\times\frac{\sin(\bar{Q}^0\ell N_{\rm rep}/2)}{\sin(\bar{Q}^0\ell/2)}\,{\cal I}_1{\rm e}^{-{\rm i}\bar{Q}^0\ell (N_{\rm rep}-1)/2}.
\label{expl}
\end{align}
We further rewrite it as
\begin{equation}
{\cal A}_{\mathrm{BW}}=\frac{\sin(\bar{Q}^0\ell N_{\rm rep}/2)}{\sin(\bar{Q}^0\ell/2)}{\cal A}_{\rm BW}^{(1)}\,{\rm e}^{-{\rm i}\bar{Q}^0\ell (N_{\rm rep}-1)/2},
\end{equation}
where ${\cal A}_{\rm BW}^{(1)}$ denotes the probability amplitude of pair creation by a single modulation of a pulse.
Introducing the parameter $N_{\rm eff}$~\eqref{neff} we arrive at
\begin{equation}
{\cal A}_{\mathrm{BW}}=\frac{\sin(\pi N_{\rm eff})}{\sin(\pi N_{\rm eff}/N_{\rm rep})}{\cal A}_{\rm BW}^{(1)}\,{\rm e}^{{\rm i}\pi N_{\rm eff}(1-1/N_{\rm rep})}.
\end{equation}
The phase factor involves the kinematic phase, $\pi N_{\rm eff}$, and the phase $\pi N_{\rm eff}/N_{\rm rep}$ which
can be interpreted as a kinematic phase resulting from a single modulation of a pulse. The latter is canceled by the
exact same phase coming from ${\cal A}_{\rm BW}^{(1)}$. The amplitude ${\cal A}_{\rm BW}^{(1)}$
carries also a dynamic phase, $\Phi_{\rm BW}^{\rm dyn}$. All together brings us to the conclusion that the
probability amplitude of the Breit-Wheeler process by a modulated laser pulse equals
\begin{equation}
{\cal A}_{\mathrm{BW}}=\frac{\sin(\pi N_{\rm eff})}{\sin(\pi N_{\rm eff}/N_{\rm rep})}|{\cal A}_{\rm BW}^{(1)}|{\rm e}^{{\rm i}\Phi_{\rm BW}},
\label{final}
\end{equation}
in agreement with Eqs.~\eqref{cp10} and~\eqref{something}.

For $N_{\rm rep}=1$ the prefactor in the above equation, which involves sine functions, equals one. For this reason, we observe very broad structures
in the energy spectra of positrons (electrons) for $N_{\rm rep}=1$. Let us define a new quantity
\begin{equation}
\frac{N_{\rm eff}}{N_{\rm rep}} = \kappa.
\label{peaks2}
\end{equation}
As long as $N_{\rm rep}>1$, the probability amplitude peaks when $\kappa$
takes only integer values, i.e., $\kappa=L=1, 2,...$. The positions of these peaks on the energy scale of positrons correspond to main maxima in Figs.~\ref{bwcomb20130322par},~\ref{bwcomb20130322per},
and~\ref{ssll2}. At these peak positions, the prefactor scales like $N_{\rm rep}$ meaning that the probability of pair creation scales like $N_{\rm rep}^2$. For this reason, the spectra
presented in the figures were divided by $N_{\rm rep}^2$; i.e., they were normalized to the spectra generated
by an umodulated pulse. Note also that under conditions when we observe the enhancement, the global phase in Eq.~\eqref{final}
becomes $\Phi_{\rm BW}=\pi LN_{\rm rep}+\Phi_{\rm BW}^{\rm dyn}$, i.e., it changes by 0 modulo $\pi$ for the subsequent values of $L$. 
This also agrees with our numerical findings (see, Fig.~\ref{bwc11}).
It follows from the formula~\eqref{final} that for $N_{\rm rep}>1$ there are $(N_{\rm rep}-1)$ zeros in-between two subsequent main peaks. Each
zero occurs at such a positron energy that $\kappa$, defined by Eq.~\eqref{peaks2}, equals $\kappa=L+M/N_{\rm rep}$ where $M=1,2,...,N_{\rm rep}-1$. Again, this is confirmed
by our numerical results. Finally, in-between each consecutive large peaks, we observe $(N_{\rm rep}-2)$ small maxima.
According to Eqs.~\eqref{final} and~\eqref{peaks2}, they occur when $\kappa=L+(M+1/2)/N_{\rm rep}$, where $M=1,2,...,N_{\rm rep}-2$. This explains
very well the patterns of positron spectra observed in Figs.~\ref{bwcomb20130322par},~\ref{bwcomb20130322per}, and~\ref{ssll2}. Finally, it follows 
from Figs.~\ref{bwcomb20130322par} and~\ref{bwcomb20130322per} that the spectra peak at the same positron energies, despite the polarization of the nonlaser photon.
This is explained by the condition $\kappa=L=1,2,...$, which does not depend on the latter. 
We would like to note that the same patterns are observed in the electron spectra as well.

\begin{figure}
\centering
\includegraphics[width=8.5cm]{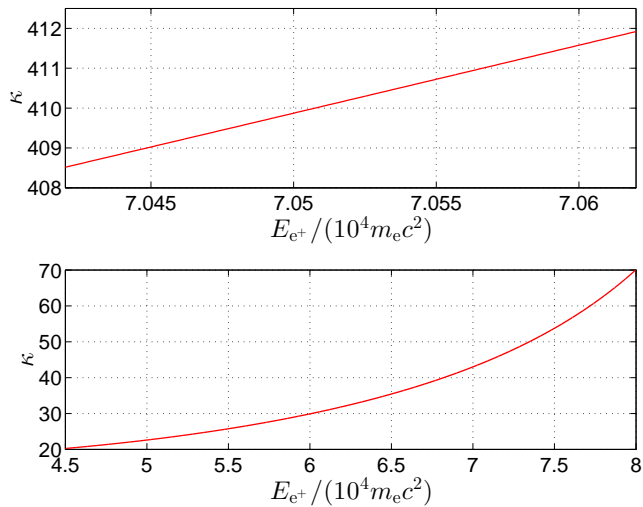}\\
\caption{The dependence of $\kappa$~\eqref{peaks2} on the positron energy. The upper panel relates to Fig.~\ref{bwc11}, whereas the lower panel
relates to Fig.~\ref{ssll2}. The same curve in each panel corresponds to three different values of $N_{\rm rep}=1,2,3$. 
At each major maximum, the parameter $\kappa$ takes an integer value. This is particularly well seen in the upper panel spanning a small
energy interval, $E_{{\rm e}^+}$, but it happens in the lower panel as well.
\label{kappa2}}
\end{figure}

We have shown that, within the finite energy interval
of product particles, equally separated and very narrow peaks in their spectra can be produced. This specifically concerns
Figs.~\ref{bwcomb20130322par} and~\ref{bwcomb20130322per}. On the other hand, we have seen from Fig.~\ref{ssll2}, which spans
a large interval of positron energy, that the enhancement peaks are distributed in a nonuniform way. This is confirmed
by plotting the dependence of $\kappa$, given by Eq.~\eqref{peaks2}, on the positron energy. In the upper panel of Fig.~\ref{kappa2}
we plot this dependence for the parameters corresponding to Fig.~\ref{bwc11}. We see that for the considered range of positron
energies, $\kappa$ passes through three integer values. This happens for the exact same positron energies for which we
observe comb maxima in the upper panel of Fig.~\ref{bwc11}. Increasing the range of positron energies such that it corresponds to Fig.~\ref{bwcomb20130322par}, 
$\kappa$ still shows the same linear dependence on $E_{{\rm e}^+}$. However, in general, $\kappa$ depends on the positron energy in a nonlinear way,
which can be observed when looking at a large enough energy interval.
This is illustrated in the lower panel of Fig.~\ref{kappa2}, which corresponds to the data collected in Fig.~\ref{ssll2}. 
We observe that with increasing the positron energy the $\kappa$-curve becomes steeper, resulting in a denser distribution of
enhancement peaks. Because the same $\kappa$ relates to modulated driving pulses with $N_{\rm rep}=2$ and $3$, we observe the enhancement peaks at the same positron energies,
independent of $N_{\rm rep}$. In fact, the same curve corresponds to an unmodulated driving pulse as well ($N_{\rm rep}=1$), since
\begin{equation}
\kappa=\frac{N_{\rm osc}}{\omega_{\rm L}}(\omega_{\bm K}-\bar{E}_{{\rm e}^+}-\bar{E}_{{\rm e}^-}).
\end{equation}
However, for the reason we have already explained, we do not observe the comb maxima for $N_{\rm rep}=1$.

\section{Conclusions}
\label{conclusions}

We have demonstrated, how by means of the nonlinear Breit-Wheeler process, one can produce coherent energy comb structures of positrons (electrons).
This requires the use of modulated laser pulses, with at least two repetitions. As shown by our numerical results, and confirmed analytically,
these energy combs are coherent. This allows for a very intuitive interpretation 
of the observed comb structures. Namely, the probability amplitudes from each subpulse interfere constructively leading to enhancements
at certain positron (electron) energies. Thus, one could say that each subpulse acts as a slit in the Young-type experiment of matter waves,
performed by Davisson and Germer \cite{Davisson}. We remark also that similar comb structures were observed in both photon and matter
domains, using nonlinear Compton scattering~\cite{KKcomb,phase}. However, the comb generation of anti-matter has been demonstrated in this
paper for the first time. As we have shown here, the comb structures extend toward GeV energies, which makes it promising for synthesis
of ultra-short bunches of anti-matter. 

In closing, we note that pair production induced by a high energy nonlaser photon takes place in a very narrow cone around the propagation
direction of the photon. In other words, the positron beam is very well collimated around the direction of the incident nonlaser photon. 
A similar situation is met in nonlinear Compton scattering, when well-collimated radiation is produced in the direction of 
energetic electron-target-beam~\cite{phase}. In both cases, i.e., for Compton scattering and
Breit-Wheeler pair production, calculations of angular distributions of product particles are very time-consuming. For this reason, in Ref.~\cite{phase}
we performed calculations using the classical theory of Compton scattering, that relates to Thomson scattering. A similar simplification 
cannot be used in the case of the Breit-Wheeler process as it has no classical equivalent.
The corresponding quantum calculations for the Breit-Wheeler process are currently being performed and their results will be published in due course.
However, based on our recent experience regarding Compton and Thomson scattering~\cite{phase}, we expect that even after integrating 
the differential positron distributions over a small solid angle, the structure of positron energy combs will still be preserved. 
Thus, we anticipate that well-resolved combs in the anti-matter domain should be detectable experimentally.
This is expected to happen due to a high collimation of positrons produced by highly energetic nonlaser photons.

\section*{Acknowledgments}
This work is supported by the Polish National Science Center (NCN) under Grant No. 2011/01/B/ST2/00381.

\end{document}